# Giant Effective Permeability in Drude Thin Films Probed by THz Time-Domain Spectroscopy

Gian Paolo Papari[1,2,*], Zahra Mazaheri[1], Antonio Vettoliere[3], Carmine Granata[3], Roberto Russo[4] , Giovanni Ausanio[1,2], Umar Farooq[1], Junaid Yaseen[1], Can Koral[5] and Antonello Andreone[1,2]

[1]Physics Department "Ettore Pancini", University of Napoli *Federico II*, Napoli, Italy

[2]CNR-SPIN, UOS Napoli, Napoli, Italy

[3]Institute of Applied Sciences and Intelligent Systems, National Research Council, Pozzuoli, Italy

[4]CNR - Institute of Applied Sciences and Intelligent Systems, Via Pietro Castellino 111, 80131 Napoli, Italy

[5]Department of Health Sciences, University of Basilicata, Potenza, Italy

Email: gianpaolo.papari@unina.it

**Abstract.**
The electromagnetic response of metallic films is commonly analyzed in terahertz spectroscopy by assuming unit relative magnetic permeability. In this work we show that this assumption introduces significant distortions in the electrodynamic retrieval of highly conductive films. Aluminum and copper films, $10nm$ thick, were investigated by terahertz time-domain spectroscopy in both transmission and reflection configurations. By applying a self-consistent retrieval method that independently determines the complex permittivity and permeability, we show that the Drude-type dielectric response is systematically accompanied by a permeability that strongly departs from unity. This deviation is intrinsically linked to the reactive impedance of the films, which clarifies the light-induced onset of large screening currents within a transversally confined geometry. A phenomenological interpretation based on the Faraday–Neumann–Lenz mechanism and a lumped-element model of the film impedance accounts for the observed trends. These results indicate that the common assumption $\tilde{\mu} = 1$ in non-magnetic Drude films can lead to an incomplete or biased electrodynamic characterization in the terahertz regime.

## 1. Introduction

The magnetic response of Drude conductors is generally considered negligible, namely

$$\tilde{\mu} = 1, \tag{1}$$

where $\tilde{\mu}$ is the relative permeability. Eq. (1) is considered to hold independently of the frequency of the external field $\overline{H}$. In practice, libraries for full-wave simulations of the electromagnetic response of common Drude conductors typically adopt the condition set by eq. (1).

In microwave [1] [2] and THz [3] bands, relevant studies made on metallic metamaterials report that full-wave simulations of permeability outside the resonance comply with the assumption that magnetic permeability is unity. In his cornerstone article [4] Veselago, while discussing the possibility of materials exhibiting both negative relative permittivity $\tilde{\varepsilon}$ and negative $\tilde{\mu}$, assumes that Drude conductors should satisfy $Re(\tilde{\mu}) > 0$ as a signature of their non-magnetic nature. In fact, the Pauli spin susceptibility for a paramagnetic free electron gas $\chi_P = \mu_0 \mu_B^2 \rho(E_F)$, where $\mu_0$ is the vacuum permeability, $\mu_B$ is the Bohr magneton and $\rho(E_F)$ is the density of states at the Fermi level, is very small for a Drude conductor ($\sim 10^{-5}$) and the Landau diamagnetism is even smaller $\chi_L = -\frac{1}{3}\chi_P$ [5].

Nevertheless, despite the intrinsically negligible atomic magnetism, the spectroscopic permeability of conducting films can deviate from eq. (1) because $\tilde{\mu}$ inherently depends on the film impedance through $\tilde{\mu} = \tilde{n}\,\tilde{z}$, where $\tilde{n}$ is the complex refractive index and $\tilde{z}$ is the impedance normalized to the vacuum impedance $Z_0 = 377\ \Omega$. Specifically, in high-conductivity films $\tilde{z} = z + i\zeta$ is expected to remain small even at high frequencies although the imaginary term $\zeta$ can increase substantially owing to the intrinsic electron-mobility-dependent kinetic inductance [5] and to the self-capacitance stemming from the closed electric field lines linked to the time-varying magnetic field $\boldsymbol{H}$ of impinging plane waves. Although small in absolute terms, the reactive contribution can produce a significant magnetic response in the frequency band from GHz to THz where macroscopic screening currents are electromagnetically induced.

Most normal-incidence spectroscopic studies on homogeneous thin metallic films focus on a single electrodynamic parameter such as the complex refractive index [6] or the complex conductivity [7] [8]. The single-parameter approach relies on the assumption $\tilde{\mu} = 1$ , that implies $\tilde{n} = 1/\tilde{z}$ thus enabling to write the Fresnel equations solely in terms of $\tilde{n}$. In reality, in order to avoid any kind of constraint in the spectroscopic retrieval, $\tilde{n}$ and $\tilde{z}$ should be measured as fully independent parameters. Investigations of these two quantities and from here $\tilde{\varepsilon} = \varepsilon_r + i\varepsilon_i$, $\tilde{\mu} = \mu_r + i\mu_i$ without any constraint have been reported for the analysis of thick samples [9] or multiferroic ($BiFeO_3$) thin films [10]. However, to the best of our knowledge, THz spectroscopy measurements to independently estimate both $\tilde{\varepsilon}$ and $\tilde{\mu}$ in homogeneous Drude metallic films have not been carried out yet.

Despite the prevailing assumptions on the permeability of Drude conductors, a rigorous retrieval process reveals that in the THz band the permeability of thin Drude-type films, such as Al and Cu, does not fulfil eq. (1). Retrieved results based on high-accuracy measurements show a large effective permeability arising from the film reactivity to THz waves.

## 2. Experimental techniques

Copper and aluminum thin films have been fabricated through sputtering techniques. Films were deposited on intrinsic Si $1 \times 2$ $cm^2$ substrates, 0.5 *mm* thick. The experiment has been designed to ensure a good signal-to-noise ratio in transmission by selecting a film thickness $t < \delta$ where $\delta = Re\{\sqrt{(2/\omega\tilde{\mu}\tilde{\sigma})}\}$ is the skin depth and $\omega$ is the angular frequency. For $f = 1\ THz$ nominal values for the skin depth are $\delta_{Al} \approx 90nm$ and $\delta_{Cu} \approx 66nm$, using bulk conductivity values $\sigma_{Al} = 3.5\ 10^7 S/m$ and $\sigma_{Cu} = 5.8\ 10^7 S/m$, for aluminum and copper respectively [11].

Atomic force microscopy (AFM) analysis of the samples reveals a surface roughness in the order of 1 *nm*. A distribution of droplets— roughly spaced 10 $\mu m$ and a few nanometers in height— is nevertheless present on the surface, a common feature in sputtered films. In Fig. 1 a representative AFM image of the Al 10 *nm* film is reported, together with a histogram showing the height distribution on the sample surface (inset).

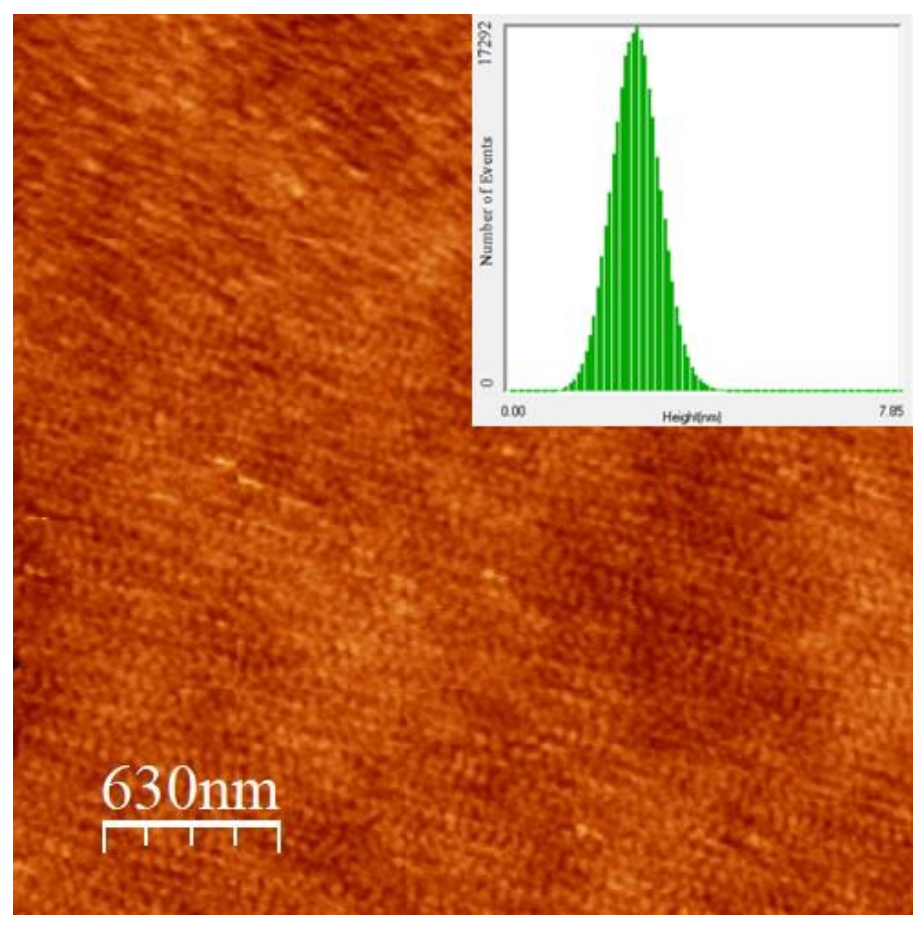

*Figure 1: AFM image of the Al film 10nm-thick. The inset shows the histogram of heights revealing that the average roughness is of the order of 1nm*

Transmission $\tilde{T}$ and reflection $\tilde{R}$ measurements were performed using a standard time domain spectrometer (TERA K-15, Menlo Systems). The two complex quantities $\tilde{T}$ and $\tilde{R}$ are acquired and processed following the same procedure described in [10]. For the bare substrate, the reference signals correspond to the free-space propagation in transmission and to a gold mirror in reflection. In contrast, for the metallic thin films the reference signals are the transmitted and reflected signals obtained from the bare Si substrate. Phase mismatch errors are minimized by acquiring the reference signal from an uncoated region of the same substrate hosting the film ("half-coated" geometry [12]). The THz beam is in fact sequentially positioned on the bare and coated areas ($1 \times 2$ $cm^2$ each) without changing alignment, ensuring that both signals share the same optical path and substrate properties. Time-dependent signals are collected in a time window of about 200 *ps* guaranteeing a frequency resolution of about 5 *GHz*. To remove unwanted water-vapor absorption, experiments were performed in a purging box filled with $N_2$ gas to keep the humidity lower than 0.1%.

Fig.2 reports the time-dependent signals corresponding to the transmission through the Cu film and the reflection from the Al, together with the respective reference signals from the Si substrate.

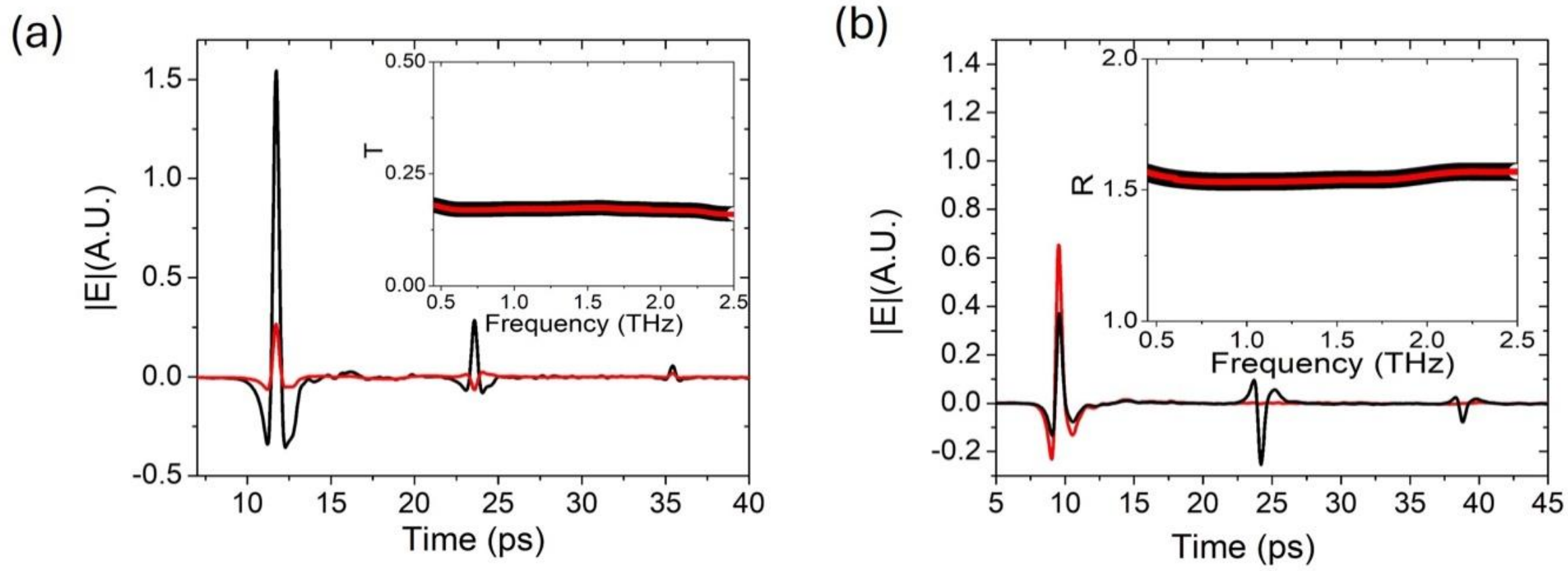

*Figure 2: (a) Black and red curves represent the time dependent signals transmitted across a Si substrate and the Cu film, respectively. Inset: Comparison between the experimental transmission (black) and the model (red). (b) Black and red curves represent the time dependent signals reflected from a Si substrate and the Al film, respectively. Inset: Comparison between the experimental reflection and the model.*

The retrieval process is a self-consistent method based on the minimization of the error calculated from the difference between the measured and the theoretical signals in frequency domain. These differences are called $Err(\tilde{T})$, $Err(\tilde{R})$ and constitute the building-blocks of the self-consistent method.

The simultaneous processing of $\tilde{T}$ and $\tilde{R}$ experimental data allows to improve the retrieval process proposed in [9] [13], enabling the development of a combined total variation technique (CTVT). This approach provides a well-defined and accurate interval in which the true value of each electrodynamic parameter lies. Specifically, the CTVT yields four optical parameters: $\tilde{n}_T$, $\tilde{n}_R$, $\tilde{z}_T$ and $\tilde{z}_R$, where the subscript denote the values obtained from the minimization of the error in transmission or reflection, respectively. Electric and magnetic responses of the material are then calculated through $\tilde{\varepsilon}_{R,T} = \tilde{n}_{R,T}/\tilde{z}_{R,T}$, $\tilde{\mu}_{R,T} = \tilde{n}_{R,T} \cdot \tilde{z}_{R,T}$.

The relative error incurred in the evaluation of electrodynamic parameters is on the order of $10^{-4}$.

The Drude-model best fits are obtained through a multivariable parameter analysis (MPA), that extends the TVT technique to the fitting routines. The Drude permittivity $\tilde{\varepsilon}_D(\omega) = 1 - \frac{\omega_p^2}{\omega^2 - i\omega\omega_\tau}$ is a function of the plasma frequency $\omega_p$, and $\omega_\tau$ is the inverse of relaxation time. The parameters $\omega_p$ and $\omega_\tau$ are determined by minimizing the error function $Err_D = \left|Re\{\tilde{\varepsilon}_D - \tilde{\varepsilon}_{exp}\}\right| + \left|Im\{\tilde{\varepsilon}_D - \tilde{\varepsilon}_{exp}\}\right|$ point by point. The MPA provides the narrowest intervals for the two independent parameters, from which average values and relative uncertainties can be extracted. It is worth noting that standard minimization techniques return the best-fit pair $\omega_p \pm \delta_{\omega_p}$, and $\omega_\tau \pm \delta_{\omega_\tau}$ so that within the intervals $[\omega_p - \delta_{\omega_p}, \omega_p + \delta_{\omega_p}]$ and $[\omega_\tau - \delta_{\omega_\tau},\ \omega_\tau + \delta_{\omega_\tau}]$ it is possible to identify the best values which minimize $\widetilde{Err_D}$ frequency by frequency. The MPA achieves the same goal, but it is specifically designed to directly minimize $\widetilde{Err_D}$ while correlating $\omega_p$, and $\omega_\tau$ at each frequency point.

### 3. The phase ranges of electrodynamic parameters

The robustness of any spectroscopic analysis relies on the mutual physical consistency of the retrieved quantities. Specifically, the reliability of the retrieved permittivity and permeability depends on the appropriate computation of $\tilde{n} = n + ik$ and $\tilde{z} = z + i\zeta$. The best way to guarantee a high quality and trustworthy retrieval is to verify that the phase ranges of $\tilde{\varepsilon}$ and $\tilde{\mu}$ remain coherently related to those of $\tilde{n}$ and $\tilde{z}$. As properly discussed in supplemental materials (SM), the available signs of electrodynamic parameters are mutually connected by the physical constraints $n, k, z > 0$. These conditions, together with the choice $\varepsilon_i > 0$, allows us to determine the ranges of $\zeta$ verifying the possible phase range of $\tilde{\mu}$. The phase ranges of $\tilde{n}$, $\tilde{z}$, $\tilde{\varepsilon}$ and $\tilde{\mu}$ are shown in Fig. 3.

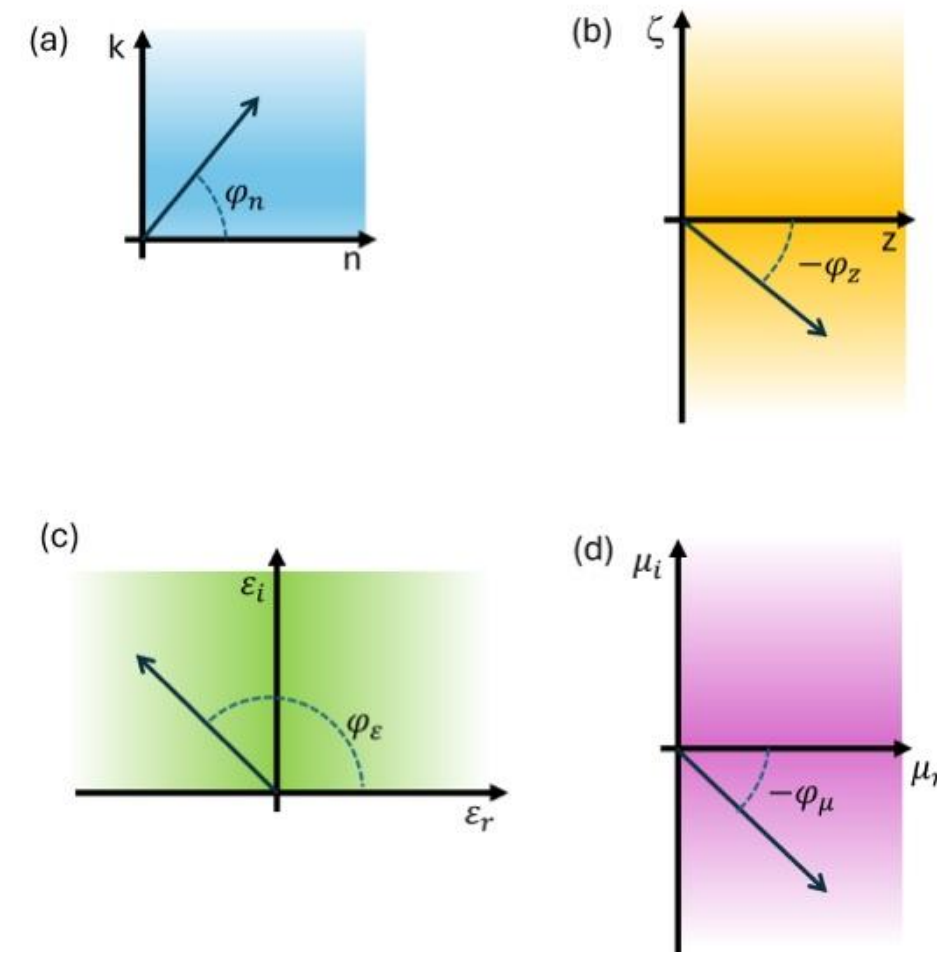

*Figure 3: Polar representations of electrodynamic parameters. The colored areas indicate the phase intervals of (a) the complex refractive index, (b) the impedance, (c) the permittivity and (d) the permeability.*

### 4. Electrodynamic parameters

An example of the high quality of the retrieval process is shown in the insets of Fig. 2 where the comparison between the experimental and modeled trends are reported.
The silicon substrate exhibits weakly dispersive electrodynamic parameters. The refractive index is mostly dominated by its real part, and the permeability is consistent with the condition $\tilde{\mu} = 1$ as expected. Further details are provided in SM. In order to highlight how misleading the assumption $\tilde{\mu} = 1$ can be in the retrieval process of highly conducting films, the comparison of the results on $\tilde{n}$ and $\tilde{\varepsilon}$ achieved by employing both the TVT (based on transmission measurements only) and the CTVT approach are presented.

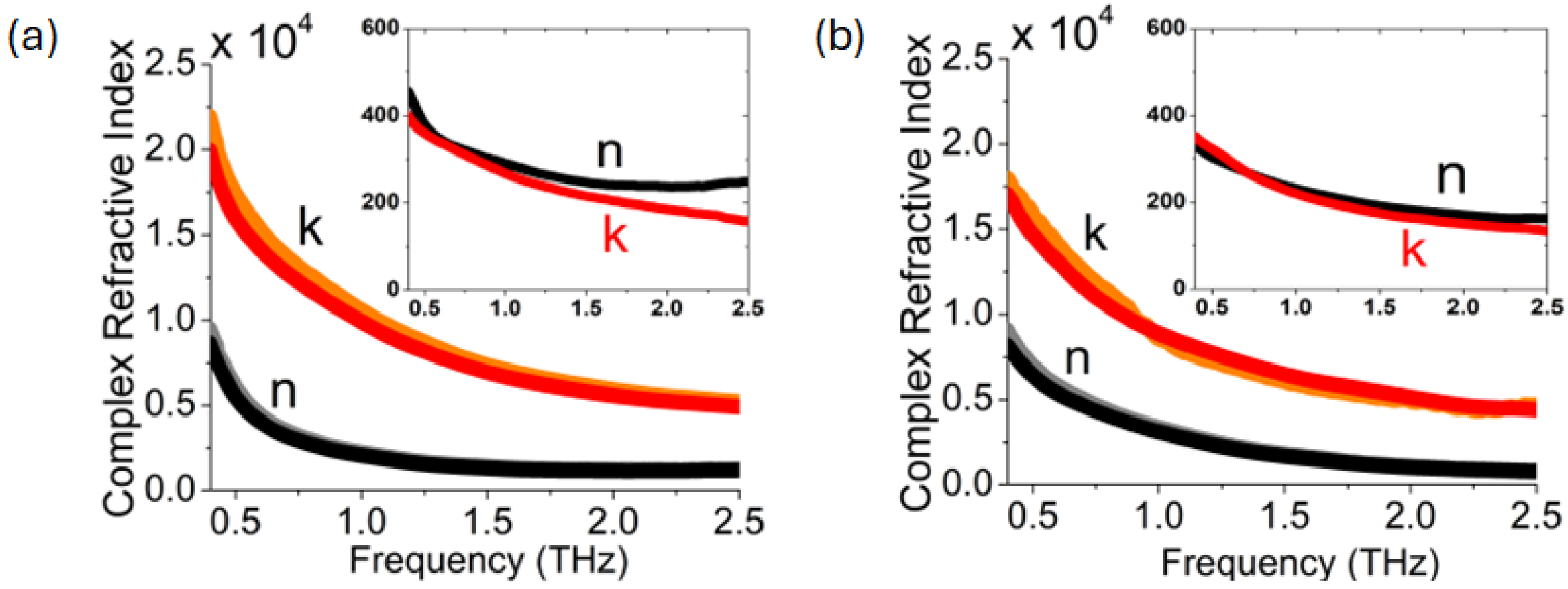

*Figure 4: In (a) and (b), the complex refractive index of the Al and Cu thin films is shown, respectively. The insets display the same parameters obtained processing the transmission measurements and applying the $\tilde{\mu} = 1$ condition.*

The retrievable spectroscopic parameters within the constraint $\tilde{\mu} = 1$ are $\tilde{n}$ and the permittivity $\tilde{\varepsilon} = \tilde{n}^2$ only [14]. In Fig. 4 the complex refractive index of Al and Cu thin films, $10\ nm$ thick, is reported. Black and red curves represent the real and imaginary parts of the electrodynamic parameters retrieved through the minimization of $Err(\tilde{T})$, whereas grey and orange curves are obtained by minimizing $Err(\tilde{R})$.

The behavior shown by both films is quite similar, presenting an extinction coefficient significantly larger than the refractive index. The metallic refractive index leads to a strong wavelength compression down to hundreds of nanometers, whereas the computation of the absorption coefficient obtained through the average extinction coefficient yields $\alpha_{avg} = \frac{2\omega k_{avg}}{c} \approx 0.5 \times 10^7 cm^{-1}$ for both metals.

The insets of Fig. 4 display the refractive indexes obtained by employing the TVT through the use of the transmission data only. In this case, the magnitude of $\tilde{n}$ is reduced by a factor $10^2$ and tends to agree with previous results achieved by using the $\tilde{\mu} = 1$ technique [6].

In Fig.5 the complex permittivity of Al and Cu films is shown. The application of the MPA routine to the permittivity of the two metals allows us to determine the values of $\omega_p, \omega_\tau$ listed in Table 1.

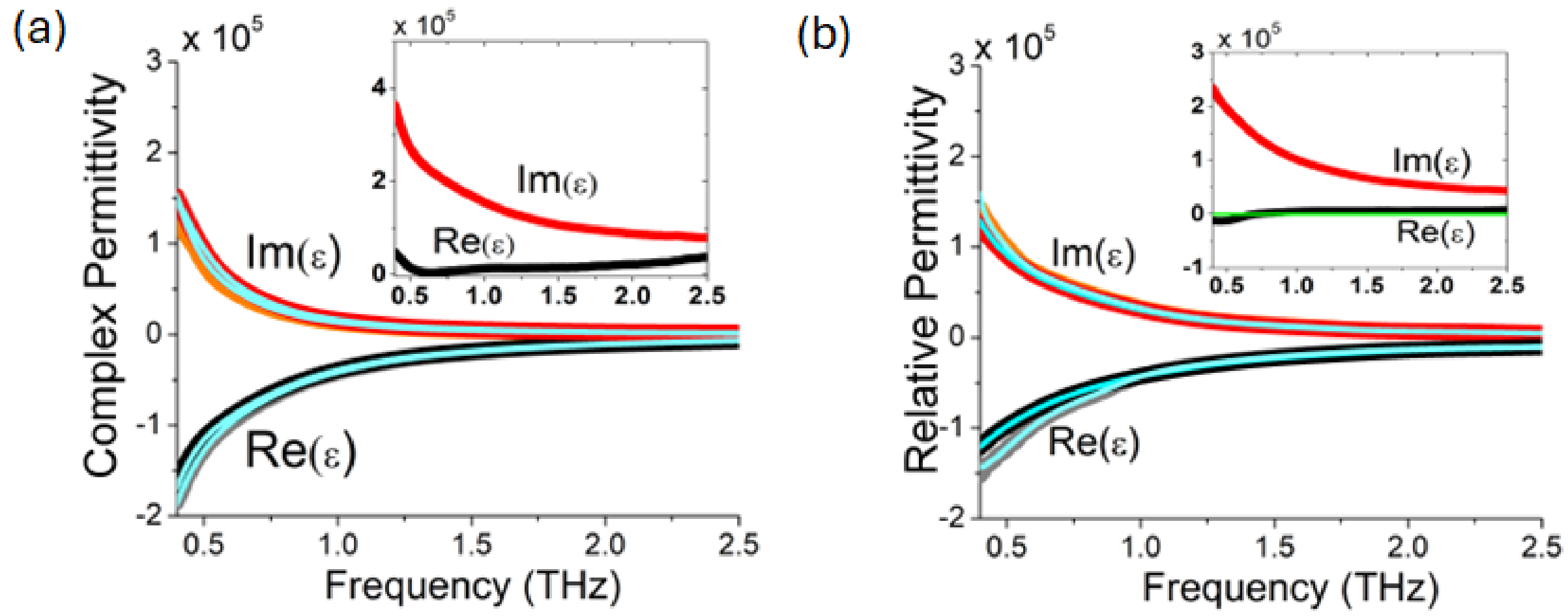

*Figure 5: In (a) and (b) the complex permittivity of the Al and Cu thin films is shown, respectively. Cyan curves represent the best-fit results obtained through the Drude model through the MPA minimization procedure. The insets display the same parameters obtained processing transmission measurements and applying the $\tilde{\mu} = 1$ condition.*

*Table 1: Best estimates of plasma and relaxation frequencies for the Al and Cu ultrathin films.*

| Metal | $\omega_p(THz)$ | $\omega_\tau(THz)$ |
|---|---|---|
| Al | $130 \pm 3$ | $2.4 \pm 0.7$ |
| Cu | $150 \pm 30$ | $4.9 \pm 2.7$ |

The best fits of such a procedure are shown in Fig. 5 as cyan curves.

The measurements reported in Table 1 allow to extract the values of the dc conductivity $\sigma_0 = \varepsilon_0 \omega_p^2/\omega_\tau$ where $\varepsilon_0$ is the vacuum permittivity. For Al and Cu the following values have been extracted: $\sigma_0^{(Al)} = (6.5 \pm 3.0)\ 10^6 S/m$ and $\sigma_0^{(Cu)} = (1.6 \pm 0.4) \cdot 10^6 S/m$, respectively. The dc conductivities are approximately one order of magnitude lower than the corresponding bulk values [11]. Certainly, sputtering does not represent the most reliable fabrication technique for producing reference films intended for the measurement of the electrodynamic parameters of "standard" materials. In fact, this route inherently induces volumetric [15] [16] and surface [17] inhomogeneities which can modify the intrinsic electron density ($n_e \cong 10^{26}/m^3$) and consequently, affect the measured conductivities of the realized specimens. Moreover, the lower conductivity observed for Cu compared with the Al one is reasonably owed to slight oxidation of the film likely due to the fact that the sample was not characterized immediately after the deposition.

In the insets of Fig. 5, the permittivity of the metallic films has been obtained through the $\tilde{\mu} = 1$ condition. Under this assumption, the Drude-like behavior is lost since $\varepsilon_r > 0$ over nearly the entire frequency range, while $\varepsilon_i$ becomes larger than the value obtained through the CTVT approach. Consequently, as reported in Fig.6, also the conductivity departs from the Drude model.

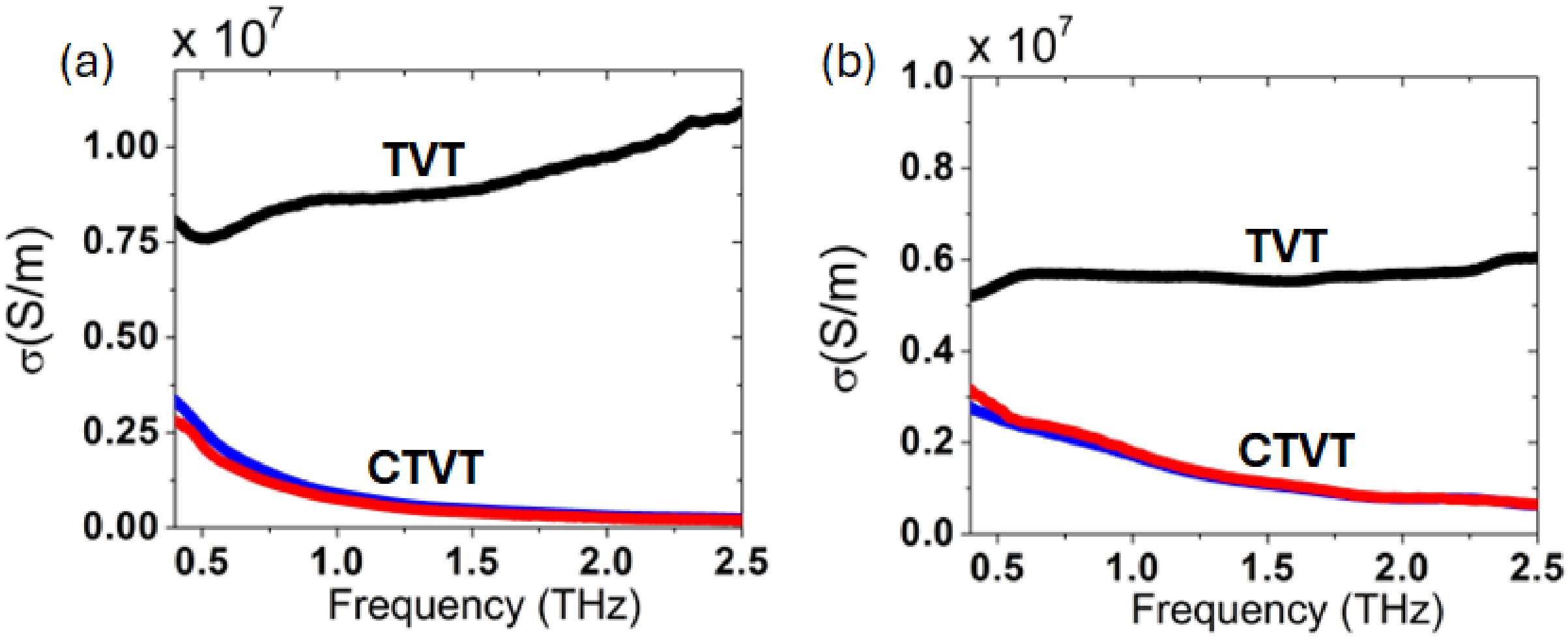

*Figure 6: The real part of conductivity for (a) Al and (b) Cu samples. Black curves represent the real part of conductivity obtained through the $\tilde{\mu} = 1$ approach (based on the TVT) whereas the blue and the red curves show $\sigma_r$ obtained through the "$\tilde{\mu} - free$" retrieval (based on the CTVT). In particular, blue and red curves have been obtained through the minimization of the error in the transmission and reflection measurements, respectively.*

Allowing the permeability to vary freely in the fitting procedure is not only more consistent from a theoretical standpoint but also leads to extracted parameters that follow the trends expected for a Drude-like material. The use of a single minimization strategy—usually based solely on the transmission data analysis and TVT retrieval [13]— yields the order of magnitude of the conductivity while loses essential information on the underlying physics. In fact, the conductivity retrieved through TVT method exhibits a rising trend for both Al and Cu films, despite the expected Drude behavior. This apparent distortion of the Drude-like conductivity has been previously observed in 10 *nm* Au thin films, where the real part $\sigma_r$ of conductivity appears to show a non-lowering behavior within a similar frequency band [7].

Under the $\tilde{\mu} = 1$ assumption, the impedance $\tilde{z}$ is the first parameter to be neglected since it collapses to $\tilde{z} = 1/\tilde{n}$. Consequently, the reactive part of the sample is artificially bound to be negative as imposed by the signs of $\tilde{n}$ components. The complex impedances of the Al and Cu films are shown in Fig. 7. As expected, the real parts are positive and small, whereas the imaginary components are negative and keep decreasing as the frequency increases. As proved below, $\tilde{z}$ is consistent with the theoretically expected behavior of the impedance of a metallic semi-infinite slab [18]:

$$\tilde{z} \cdot Z_0 = (1 - i)\sqrt{\frac{\omega\mu_0\tilde{\mu}}{2\tilde{\sigma}}} = \sqrt{-i\,\frac{\omega\mu_0\tilde{\mu}}{\tilde{\sigma}}}. \quad (2)$$

This indicates that its physical origin can be attributed to the macroscopic behavior of the induced currents involving the whole sample. As a consequence of the inequality $\lambda_{in} \gg t$, where $\lambda_{in} = \lambda/n$, the dynamics of impedances can be described through a lumped-element model. The impedance of a metallic film can be modeled through the parallel of two branches: one is resistive-inductive whereas the second is a capacitive impedance [19] [20]. The RL branch accounts for the plasma behavior of electrons under the effects of $\boldsymbol{E}$ while the (self) capacitive impedance arises from $\boldsymbol{E}_H$, the electric field linked to $\boldsymbol{H}(t)$, related to displacement currents defined by closed lines linking two points of the film across air. Denoting by $R$, $L$ and $C$ respectively the effective resistance, inductance and capacitance of the film, the equivalent impedance is modeled as

$$\tilde{z} \cdot Z_0 = \tilde{Z}_{eq}(\omega) = \frac{R + i\,\omega L(\omega)}{1 - \omega^2 L(\omega)C(\omega) + i\,\omega C(\omega)}. \quad (3)$$

By assuming $R = 1/\sigma_0 t$ and according to the range of $\sigma_0$, MPA was used to obtain the best couple $L(\omega), C(\omega)$ for each frequency point. Since for each sample $\tilde{z}_T$ and $\tilde{z}_R$ nearly overlap, we have performed the MPA on the average impedance. The best-fit results of $\tilde{Z}_{eq}/z_0$ are reported in Fig. 7 through green (real part) and yellow (imaginary part) curves.

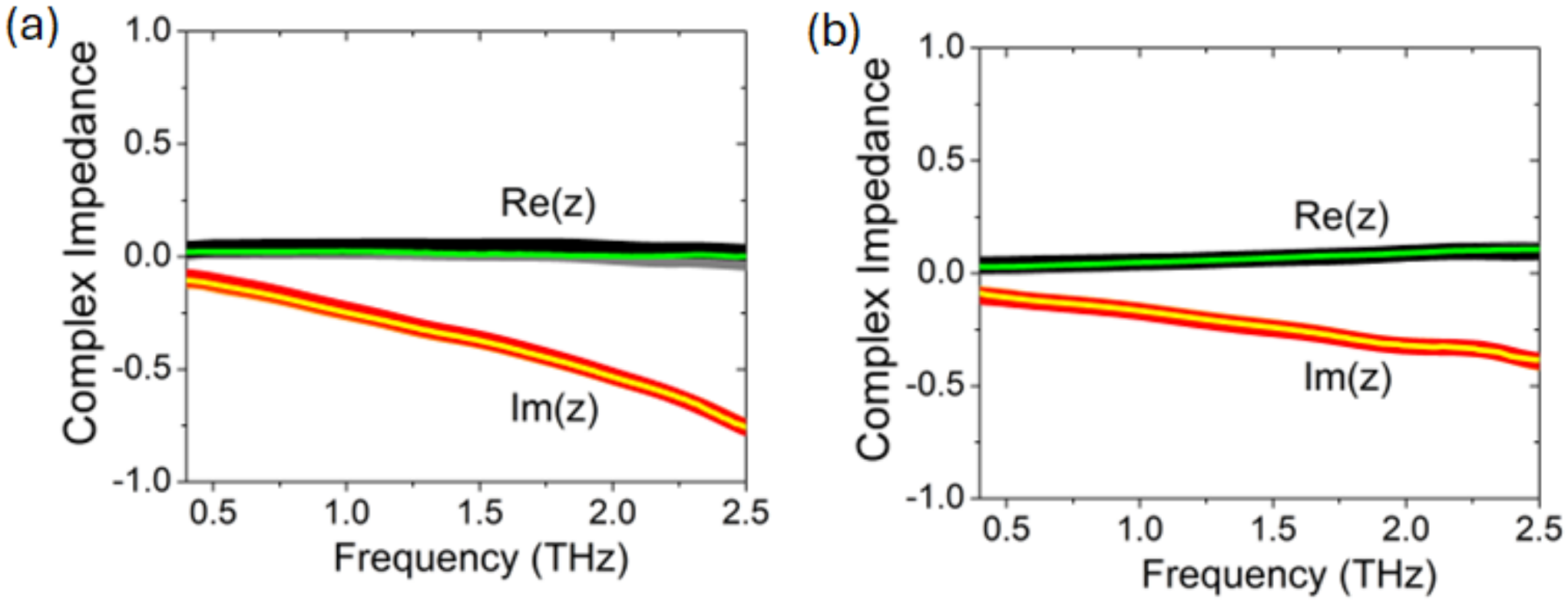

*Figure 7: In (a) and (b) the complex impedance of the Al and Cu thin films is reported, respectively. Green and yellow curves represent the MPA-based best fits according to eq. (3). See the text for further details.*

The achieved intervals of inductance and capacitance for the aluminum and copper films are $L_{Al} \in [13, 40]\ pH$, $C_{Al} \in [0.4, 16]\ fF$ and $L_{Cu} \in [12, 23]\ pH$ , $C_{Cu} \in [0.70, 13]\ fF$ respectively. The retrieved inductances range in an interval confident with the kinetic inductance $L_k = \frac{m}{n_e e^2}\frac{l}{wt} = \left(\frac{\tau}{\sigma_0}\right)\frac{l}{wt}$, where $l, w$ and $t$ represent the length (parallel to $\boldsymbol{E}$), the width (perpendicular to $\boldsymbol{E}$) and the thickness of a metallic slab, respectively. For $l =$ $w$ (unit surface area) we obtain the kinetic inductance per square which, according to the retrieved ranges of relaxation time ($\tau = \frac{1}{\omega_\tau}$) and dc conductivity, lies in the following intervals: $L_k^{(Al)} = [4.0, 40]\ pH$ and $L_k^{(Cu)} = [7.0, 30]\ pH$. The retrieved values for $C$ are also consistent with the expected self-capacitance $C_s \approx 2\pi\ \varepsilon_0\ l$ [21]. By assuming as reference length $l = [10.0, 750]\ \mu m$, which accounts for the defects (droplets) main distance and the range of the impinging wavelengths, the values of self-capacitance range in $C_s \in [0.50, 30] fF$. Thus, both values of retrieved $L$ and $C$ agree with the possible quantities of kinetic inductance and self-capacitance.

Once the changes in $\tilde{n}$, $\tilde{\varepsilon}$ and $\tilde{\sigma}$ induced by the $\tilde{\mu} = 1$ assumption have been clarified, the large values obtained for the retrieved permeabilities should no longer be viewed as unexpected. Fig. 8 shows the extracted giant effective permeability of Al and Cu thin film samples.

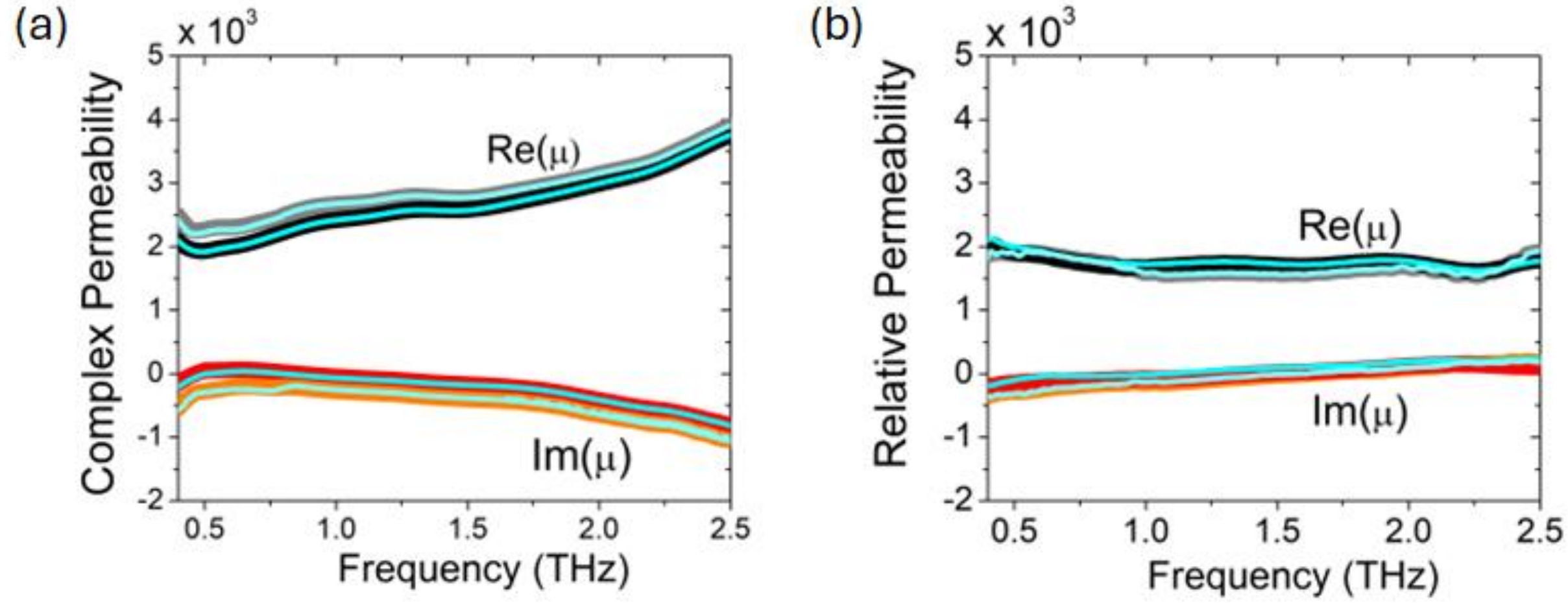

*Figure 8: (a) and (b) report the complex permeability of the AL and Cu thin films, respectively. Black, red, grey and orange curves represent the measurements according to the CTVT whereas the cyan curves have been obtained by using eq. (4).*

The extremely large values of permeability cannot be ascribed to intrinsic magnetism of Drude conductors which, from the single-electron standpoint, react to an external magnetic field only through the spins of a free electron gas [5]. However, before debating on the origin of the giant permeability, it is important to verify that the measured $\tilde{\mu}$ is coherently connected to the other electrodynamic parameters. By exploiting eq. (2), the permeability can be expressed in terms of the measured $\tilde{\sigma}$ and $\tilde{z}$ yielding

$$\tilde{\mu} = i\tilde{\sigma}\ Z_0^2 \tilde{z}^2 / \omega\ \mu_0. \tag{4}$$

If in the latter equation the measured values of the conductivity and impedance are used, the cyan curves in Fig. 8 are obtained.

According to eq. (4), the verified inequality $|\tilde{\mu}_{Al}| > |\tilde{\mu}_{Cu}|$ is most likely consequence of the higher conductivity measured for Al compared to Cu, given that $\tilde{z}_{Al} \sim \tilde{z}_{Cu}$.

Furthermore, the Al film reveals an apparent diamagnetic behavior ($\mu_i < 0$) whereas the Cu-film permeability shows a weak dissipation in the investigated band. In order to provide a clearer description of $\tilde{\mu}$ in thin metallic films, it is necessary to gain further insight into the development of screening currents throughout the film.

## 5. The skin depth

Within the metal, $\boldsymbol{E}$ decays according to the law $E(x, \omega) = E_0\, e^{-x/\delta(\omega)}$, where $E_0$ represents the field at the interface and $\delta(\omega) = Re\left\{\sqrt{2/\omega\tilde{\sigma}\mu_0\tilde{\mu}}\right\}$ is the skin depth [18]. If screening currents are active within the sample, the skin depth is defined through the retrieved $\tilde{\mu}$, otherwise the skin depth should be calculated by setting $\tilde{\mu} = 1$. In Figs. 9(a) and (b) the skin depth is reported for the case $\tilde{\mu}$ *−free* (black and red curves) and compared with the spectroscopic results obtained using the transmission only (green curve). In the latter case the skin-depth is larger than a factor 10 with respect the $\tilde{\mu}$ *−free* case. The transmission only spectroscopy provides a $\delta(\omega)$ value that does not represent the observed phenomenology because the correspondence $|\tilde{T}| \sim e^{-t/\delta}$ is not verified. The validity of the skin depth retrieved through the CTVT is observable in Figs. 9(c) and (d), where the exponential decay is compared with the measured transmission represented by magenta dots. The agreement between the exponential decay defined by $\delta_{CTVT}$ ensures that the magnetic coupling is preserved within the metal thickness as well.

The outcomes on the skin depth suggest that the origin of the giant permeability stems from the development of screening currents throughout the films. The origin of the large $\tilde{\mu}$ can be ascribed to the development of screening currents linked to the time-varying flux of the magnetic field getting across the film.

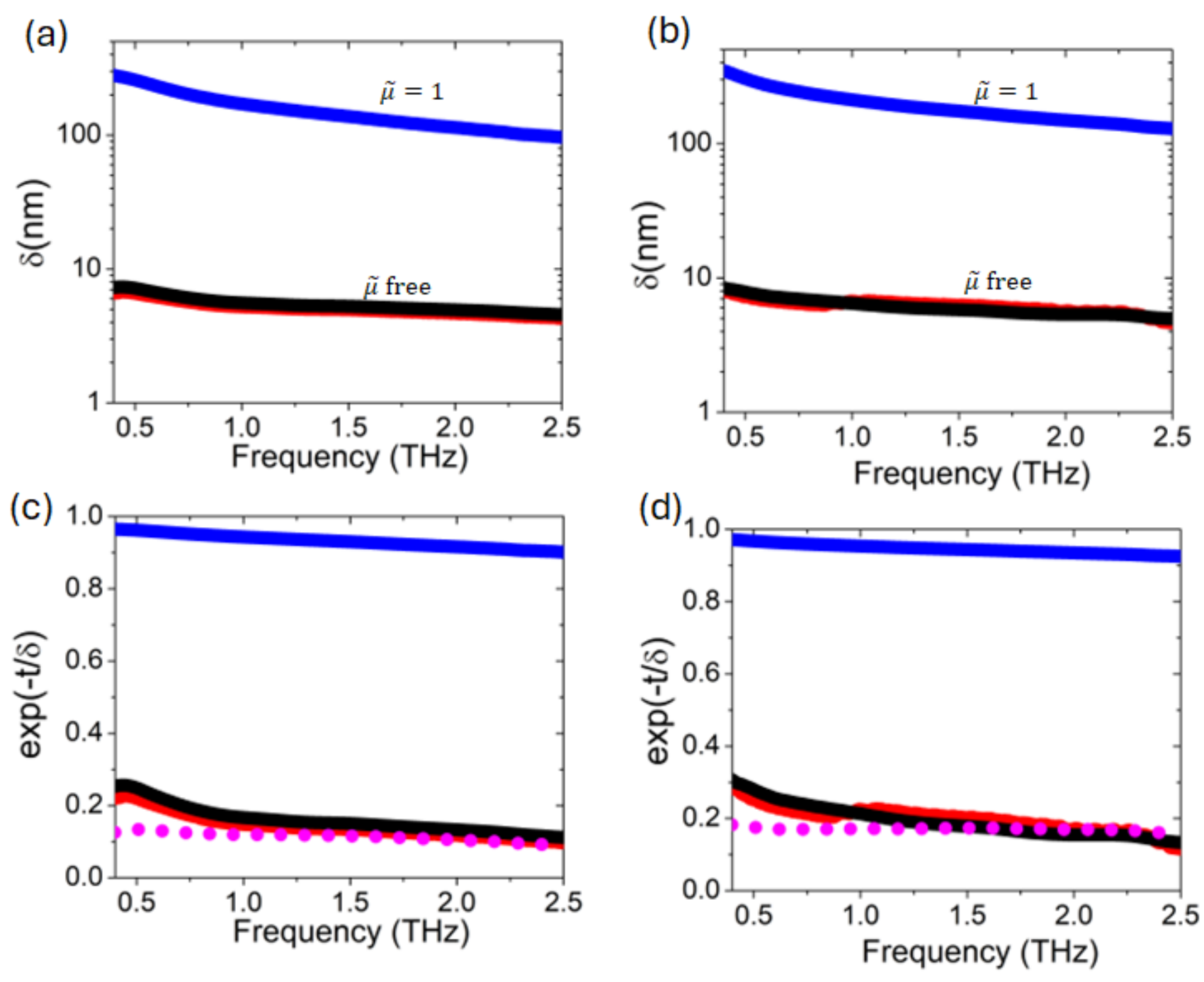

*Figure 9. In (a) and (b) the skin depths of the Al and the Cu films are reported, respectively. Black and red curves represent the skin depths obtained through the CTVT whereas the blue curve is the skin depth measured through the TVT. In (c) and (d) the exponential decay defined by the skin depth of Al and Cu is compared with the measured transmission. Black and red curves have been obtained using the CTVT whereas the blue one through the TVT. The measured transmissions are reported as magenta dots.*

## 6. Origin of the effective magnetic response

Experimental outcomes indicate that the extremely high values of $\tilde{\mu}$ are originated by the involvement of large currents which can be phenomenologically linked to the Faraday–Neumann–Lenz (FNL) effect. The onset of FNL contribution arises from the time-varying magnetic flux occurring in the interaction between the THz wave and the thin metallic film. The impedance of a metallic film is usually assumed to be dependent on $\tilde{\mu} = 1$ [18] [22], but we examine that the impedance is connected to the work on charge carriers of the electric field $\boldsymbol{E}_H(t)$ linked to the time-varying magnetic flux.
The key element is the lumped circuit $\tilde{z}$, depicted in Fig.10, showing the $RL$ representing the Drude response and the $C$ branch that bridges —through air— two points of the metal surface. According to the discussions on $\lambda_{in} \gg t$ and the skin depth, the lumped circuit describes the behavior of both surface and volumetric currents.

The phenomenological interpretation of the FNL contribution is based on the reconstruction of the impedance expression (eq. 2), starting by the time-varying magnetic flux across the lumped circuit.

The FNL effect is described by the equation $\Delta V_H = -d\Phi/dt$, where $\Delta V_H = \oint \boldsymbol{E}_H \cdot d\boldsymbol{l}$ is the voltage drop along the closed circuit and $\boldsymbol{E}_H$ is the electric field linked to $H$. By extending the latter equation to the lumped circuit, the flux across the surface ($S$) delimited by the capacitive and the resistive-inductive branch can be calculated. The voltage drop across the circuit is given by the sum of the currents flowing through the $RL$ branch ($I_D$) and the capacitive one (the displacement current $I_C$). Thus, the FNL equation turns into

$$\tilde{z}_f(I_D + I_C) = -i\omega\tilde{\mu}HS. \quad (5)$$

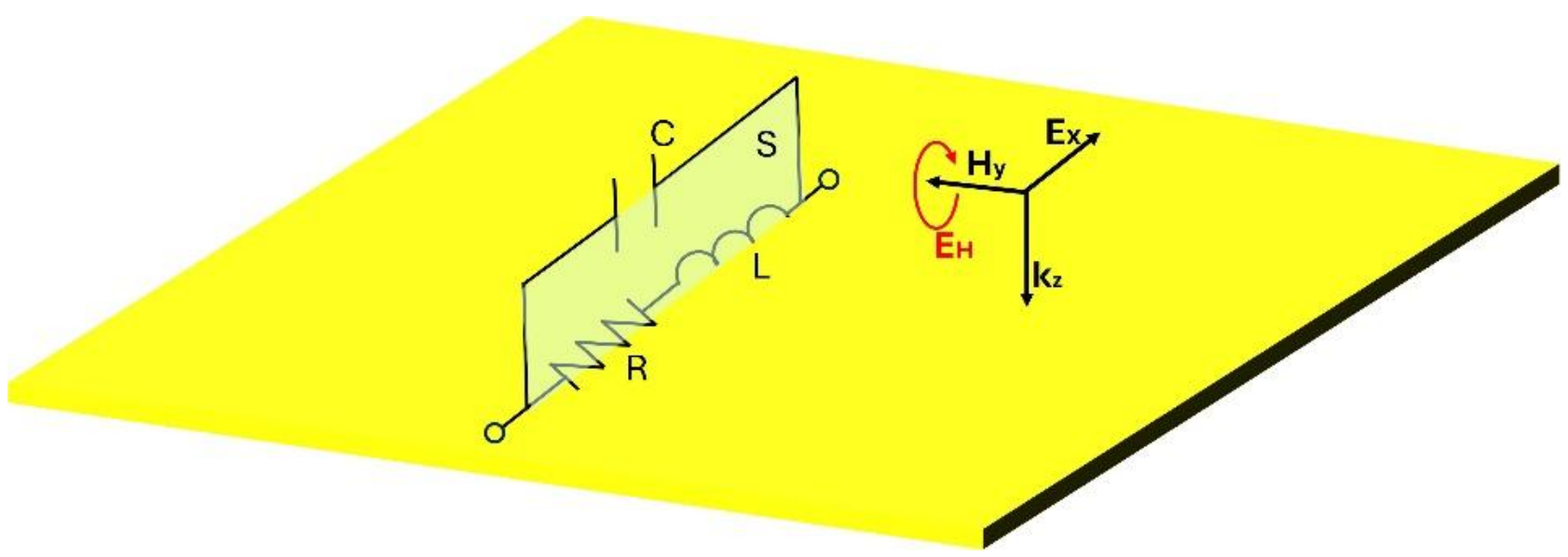

*Figure 10: Schematic representation of the lumped impedance defining the electromagnetic coupling with the film.*

From $I_C\tilde{z}_C = I_D\tilde{z}_D$ it is clear that $I_C \ll I_D$ holds true, since $\tilde{z}_C \gg \tilde{z}_D$. Thus, the left member of eq. (5) can be expressed in terms of $I_D$ only.

We now pass from the macroscopic view to the microscopic one by specifying the geometrical dimensions of the circuit. This has a size $l \times w \times T$, where $l$ is the effective circuit length ($\parallel \boldsymbol{E}$), $w$ is its width ($\parallel \boldsymbol{H}$) and $d$ represents the circuit height ($\parallel \boldsymbol{k}$) proportional to $\lambda$ since referring to the extension of $\boldsymbol{E}_H$ lines. Thanks to the following substitutions: $S = l \cdot d$, $I_D = J_D\, w \cdot d$, $\tilde{z} = E/H$, $J_D = \tilde{\sigma}\, E$, from eq. (5) the permeability can be expressed as $\tilde{\mu} = i\tilde{z}^2\tilde{\sigma}\, w/l\, \omega$. Since the size of the lumped circuit can be assumed as proportional to the wavelength, namely $w = l \propto \lambda$, the previous expression recovers the same form of the permeability reported in eq. (4). Hence, the consistency between $\tilde{\mu}$ and $\tilde{z}$ is supported by the presence of the capacitive branch "activated" by $\boldsymbol{H}(t)$. Along $k_z$ the THz waves, characterized by $(\lambda, \lambda_{in}) \gg t$, sense a single circuit that disables any possibility of screening currents cancellation between adjacent loops. This is the reason behind the onset of the giant permeability. The $\tilde{\mu} \neq 1$ phenomenon occurs every time the inequality $\lambda_{in} \gg t$ is satisfied, provided that $\delta \geq t$ to allow the transmission and consequently the measure of $\tilde{\mu}$. However, as long as the following inequality is satisfied

$$\lambda_{in} > \delta, \quad (6)$$

the screening currents cannot undergo self-cancellation and the phenomenon $\tilde{\mu} \neq 1$ occurs even if the sample is so thick to present $\tilde{T} = 0$. In fact, the measured films behave as quasi-infinite slabs featured by $\coth(\tilde{\gamma}t) \approx 1$, where $\gamma = \sqrt{i\omega\mu_0\tilde{\mu}\tilde{\sigma}}$ is the propagation factor. The hyperbolic cotangent factor accompanies the impedance (eq. 2) only when the sample is optically thin. In addition, the measurements yield $|\tilde{T}| \approx 0.1$, indicating a behavior more consistent with a thick rather than a thin sample.
A pictorial representation of the situations $\lambda_{in} > \delta$ and $\lambda_{in} < \delta$ is reported in Fig. 11.

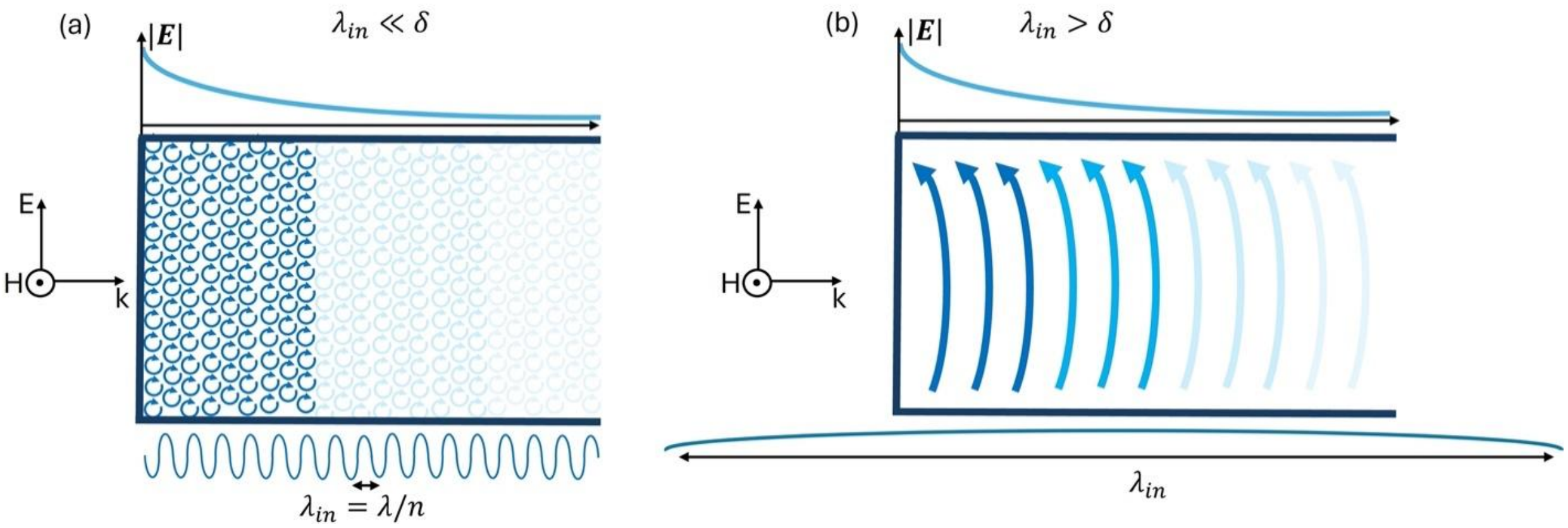

*Figure 11. (a) Schematic representation of screening current activation for an optically thick conducting sample in the case $\lambda_{in} \ll \delta$. In this case the overall contribution of screening currents is zero and $\tilde{\mu} \to 1$. (b) Illustrative depiction of the non-null screening currents contribution relative to the case $\lambda_{in} > \delta$.*

Furthermore, the dependence of $\tilde{\mu}$ on the activation of screening currents indicates that the para- or the dia-magnetic behavior of the sample is certainly does not rely on the atomic species, but it derives from a macroscopic behavior of $\boldsymbol{H}$(t)-induced screening currents. In this respect, the film thickness and its morphology over the scale of $\lambda$ is likely to affect the sign of $\mu_i$ as well as the amplitude of $\mu_r$. The para- or the dia-magnetic response can be rationalized as the dissipation induced by $\boldsymbol{H}(t)$ is in phase or in counter-phase ($\pi$ shifted) with respect to the one induced by $\boldsymbol{E}(t)$, namely $\varepsilon_i$.

## 7. Conclusions

An accurate self-consistent method has been applied to retrieve the electrodynamic parameters of two Al and Cu thin films in the THz band. The present investigation indicates that, to avoid misinterpretation in the electrodynamic retrieval, it is essential to investigate the electrodynamic properties of conductive materials by performing the independent acquisition of $\tilde{n}$ and $\tilde{z}$. Combined THz transmission–reflection retrieval reveals that thin Drude films cannot be consistently described by imposing $\mu_r$ =1 within the adopted effective-medium model.

The possible signs of $\tilde{\mu}$ components have been addressed through the mutual consistency of electrodynamic parameters. $\mu_r$ is always positive whereas the sign of $\mu_i$ configures the dia- or para-magnetic behavior of the sample. The retrieved permeability is extremely large for both the investigated films, signaling the onset of a neglected mechanism in the spectroscopy of conducting films. The giant effective permeability is interpreted as the signature of the FNL effect occurring across the film. Phenomenologically, the effect can be described by considering the time-varying flux of the magnetic field $\boldsymbol{H}(t)$ across the film whose impedance is modeled through a single lumped-element circuit, composed by the parallel of a Drude (RL) branch and a capacitive branch. The former branch motivates the Drude behavior, whereas the closed loop justifies the activation of screening currents induced by electric field lines linked to $\boldsymbol{H}(t)$. Within this framework, the meaning of $\tilde{\mu}$ is not related to any intrinsic magnetic property, instead it stems entirely from the induced screening currents developed within the thin layer preventing their self-cancellation. Since the phenomenon occurs when the inequality $\lambda_{in} > \delta$ is verified, the giant permeability phenomenon should rise also in thick samples where the transmission and consequently $\tilde{\mu}$ cannot be measured.

The condition $\tilde{\mu} \neq 1$ in conducting materials strongly affects multiple aspects of matter-radiation phenomena. For instance, the coherent spectroscopy of conducting and/or magnetic materials, the electrodynamics of both diluted metals and metasurfaces all require a thorough re-evaluation in light of the present results.

**Acknowledgments**

We are grateful to Prof. Vladimir Fomin for his helpful suggestions that helped refine the text.

**Data Availability Statement**

The data are available upon reasonable request.

**Author Contributions**

G.P. designed the experiment, processed the data, conceptualized the manuscript and wrote the initial draft of the manuscript. Z.M. and C.K. performed the experiments, A.V., C.G., R.R. and G.A. fabricated the samples and carried out the morphological characterization. U.F. and J.Y. verified the processed data and prepared the figures. A.A.

supported data interpretation and refined the manuscript. All authors discussed the results and approved the final version of the manuscript.

**References**

[1] D. R. Smith, D. C. Vier, T. Koschny, and C. M. Soukoulis, Electromagnetic parameter retrieval from inhomogeneous metamaterials, Phys. Rev. E **71**, 036617 (2005).

[2] J. Zhou, L. Zhang, G. Tuttle, T. Koschny, and C. M. Soukoulis, Negative index materials using simple short wire pairs, Phys. Rev. B **73**, 041101(R) (2006).

[3] S. Zhang, Y. S. Park, J. Li, X. Lu, W. Zhang, and X. Zhang, Negative refractive index in chiral metamaterials, Phys. Rev. Lett. **102**, 023901 (2009).

[4] V. G. Veselago, The electrodynamics of substances with simultaneously negative values of ε and μ, Soviet Physics Uspekhi **10**, 509 (1968).

[5] N. W. Ashcroft and N. D. Mermin, *Solid State Physics* (New York, USA, 1976).

[6] D. X. Zhou, E. P. J. Parrott, D. J. Paul, and J. A. Zeitler, Determination of complex refractive index of thin metal films from terahertz time-domain spectroscopy, J. Appl. Phys. **104**, 053110 (2008).

[7] M. Walther, D. G. Cooke, C. Sherstan, M. Hajar, M. R. Freeman, and F. A. Hegmann, Terahertz conductivity of thin gold films at the metal-insulator percolation transition, Phys. Rev. B Cond. Matt. Phys. **76**, 1 (2007).

[8] N. Laman and D. Grischkowsky, Terahertz conductivity of thin metal films Terahertz conductivity of thin metal films, Appl. Phys. Lett. **93**, 051105 (2008).

[9] H. Němec, F. Kadlec, P. Kužel, L. Duvillaret, and J. L. Coutaz, Independent determination of the complex refractive index and wave impedance by time-domain terahertz spectroscopy, Opt. Commun. **260**, 175 (2006).

[10] G. P. Papari, Z. Mazaheri, F. Lo Presti, G. Malandrino, and A. Andreone, Accurate THz measurements of permittivity and permeability of $BiFeO_3$ thin films, Opt. Commun. **586**, 131872 (2025).

[11] C. Kittel, *Introduction to Solid State Physics*, 8th Edition (Hoboken, NJ, 2005).

[12] F. Yan, E. P. J. Parrott, B. S. Y. Ung, and E. Pickwell-Macpherson, Solvent doping of PEDOT/PSS: Effect on terahertz optoelectronic properties and utilization in terahertz devices, Journal of Physical Chemistry C **119**, 6813 (2015).

[13] L. Duvillaret, F. Garet, and J. L. Coutaz, A reliable method for extraction of material parameters in terahertz time-domain spectroscopy, IEEE Journal on Selected Topics in Quantum Electronics **2**, 739 (1996).

[14] A. Pashkin, M. Kempa, H. Němec, F. Kadlec, and P. Kužel, Phase-sensitive time-domain terahertz reflection spectroscopy, Review of Scientific Instruments **74**, 4711 (2003).

[15] M. Samuelsson, D. Lundin, J. Jensen, M. A. Raadu, J. T. Gudmundsson, and U. Helmersson, On the film density using high power impulse magnetron sputtering, Surf. Coat. Technol. **205**, 591 (2010).

[16] M. Draissia, H. Boudemagh, and M. Y. Debili, Structure and hardness of the sputtered Al-Cu thin films system, Phys. Scr. **69**, 348 (2004).

[17] G. Zhu, M. Han, B. Xiao, and Z. Gan, On the microcrystal structure of sputtered Cu films deposited on Si(100) surfaces: experiment and integrated multiscale simulation, Molecules **28**, (2023).

[18] R. E. . Collin, *Foundations for Microwave Engineering*, second edition (Wiley-IEEE Press, New York, NY, USA, 2015).

[19] G. P. Papari, C. Koral, and A. Andreone, Geometrical dependence on the onset of surface plasmon polaritons in THz grid metasurfaces, Sci. Rep. **9**, 924 (2019).

[20] K. Takano, K. Shibuya, K. Akiyama, T. Nagashima, F. Miyamaru, and M. Hangyo, A metal-to-insulator transition in cut-wire-grid metamaterials in the terahertz region, J. Appl. Phys. **107**, (2010).

[21] Simon. Ramo, J. R. . Whinnery, and Theodore. Van Duzer, *Fields and Waves in Communication Electronics* (Wiley, 1994).

[22] S. A. Maier, *Plasmonics: Fundamentals and Applications* (Springer US, New York, NY, 2007).

Supplemental material

# Giant Effective Permeability in Drude Thin Films Probed by THz Time-Domain Spectroscopy

Gian Paolo Papari[1,2,*], Zahra Mazaheri[1], Antonio Vettoliere[3], Carmine Granata[3], Roberto Russo[4] , Giovanni Ausanio[1,2], Umar Farooq[1], Junaid Yaseen[1], Can Koral[5] and Antonello Andreone[1,2]

[1]Physics Department "Ettore Pancini", University of Napoli "Federico II", Napoli, Italy

[2]CNR-SPIN, UOS Napoli, Napoli, Italy

[3]Institute of Applied Sciences and Intelligent Systems, National Research Council, Pozzuoli, Italy

[4]CNR - Institute of Applied Sciences and Intelligent Systems, Via Pietro Castellino 111, 80131 Napoli, Italy

[5]Department of Health Sciences, University of Basilicata, Potenza, Italy

## 1. Discussion on the phase ranges of electrodynamic parameters

In a previous publication [1], the negative values of $\mu_i$ have been demonstrated to be physically acceptable. In the following section, the phase range of $\tilde{\mu}$ is rigorously discussed together with the possible signs of real and imaginary components of $\tilde{n}$, $\tilde{z}$ and $\tilde{\varepsilon}$.

The polar representations of the complex refractive index $\tilde{n} = n + ik$ and the impedance $\tilde{z} = z + i\zeta$ are given by the following equations

$$\tilde{n} = |\tilde{\varepsilon}\tilde{\mu}|e^{i(\varphi_\varepsilon+\varphi_\mu)/2} = |\tilde{n}|e^{i\varphi_n} \tag{1}$$

$$\tilde{z} = |\tilde{\mu}/\tilde{\varepsilon}|e^{i(\varphi_\mu-\varphi_\varepsilon)/2} = |\tilde{z}|e^{i\varphi_z} \tag{2}$$

where $\varphi_\varepsilon = \arctan \varepsilon_i/\varepsilon_r$ and $\varphi_\mu = \arctan \mu_i/\mu_r$ denote the phases of the permittivity and permeability, whereas $\varphi_n$ and $\varphi_z$ represent the phases of the complex refractive index and impedance, respectively.

Among the components defining $\tilde{n}$ and $\tilde{z}$, only $\zeta$ can take positive and negative values whereas the others must satisfy

$$\begin{cases} n > 0 \\ k > 0 \\ z > 0 \end{cases} . \tag{3}$$

These constraints imply that the phase ranges of $\tilde{n}$ and $\tilde{z}$ must satisfy $\varphi_n \in [0, \pi/2\,]$ and $\varphi_z \in [-\pi/2\,, \pi/2\,]$ respectively. These phase ranges are illustrated in Figs. S1(a) and (b).

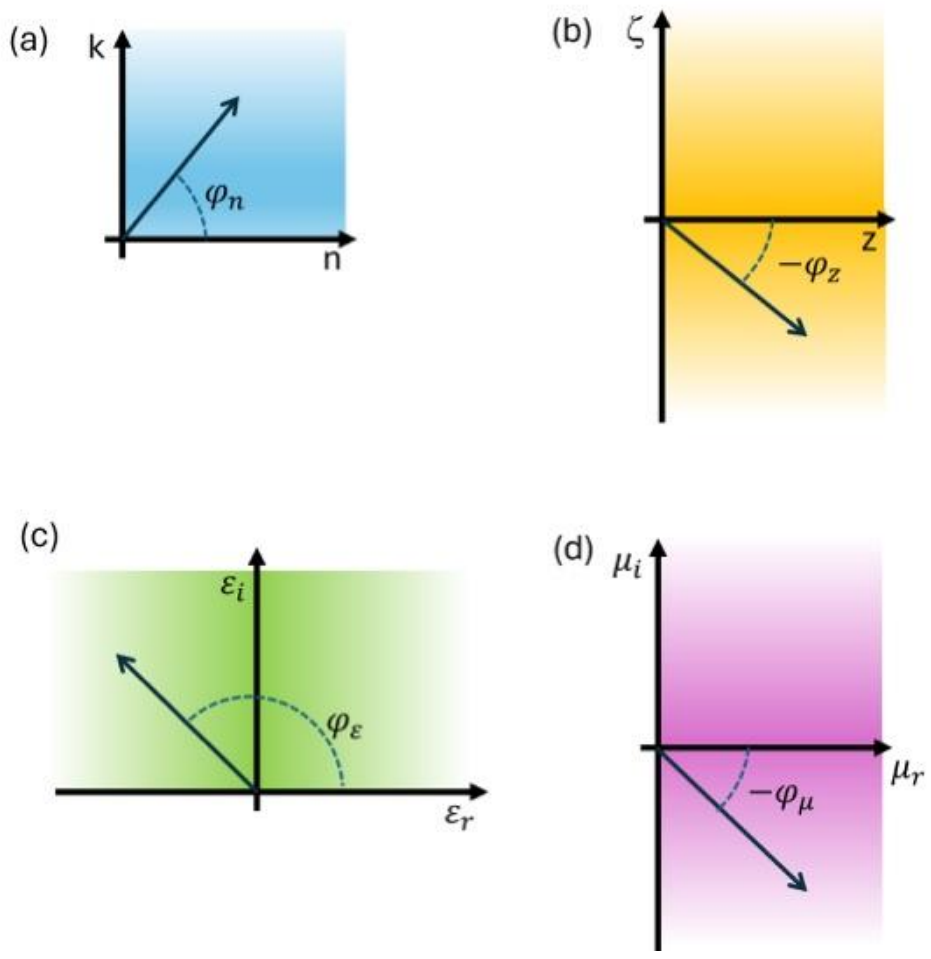

*Figure S1: Polar representations of electrodynamic parameters. The colored areas indicate the phase intervals of (a) the complex refractive index, (b) the impedance, (c) the permittivity and (d) the permeability.*

The phase range of permittivity is reported in Fig. S1 (c) and satisfies $\varphi_\varepsilon \in [0, \pi]$. The available phase range of $\tilde{\mu}$ is commonly assumed to match with the permittivity one, namely $\varphi_\mu \in [0, \pi]$. In the following, we show that the phase range of permeability, to remain physically consistent with phase intervals of the other electrodynamic parameters, must be $\varphi_\mu \in [-\pi/2, \pi/2]$ as reported in Fig. S1(d).

The four possible sign combinations of $\tilde{\mu}$ components are: *(i)* $\mu_r > 0$, $\mu_i > 0$; *(ii)* $\mu_r > 0$, $\mu_i < 0$; *(iii)* $\mu_r < 0$, $\mu_i > 0$; *(iv)* $\mu_r < 0$, $\mu_i < 0$. Since $\mu_r$ and $\mu_i$ may be expressed as a function of $n, k, z$ and $\zeta$, the constraints reported in (3) may be used to determine the ranges of $\zeta$ for which the regimes *(i)-(iv)* may occur. Specifically, regime *(i)* is satisfied for $-\frac{k}{n}z < \zeta < \frac{n}{k}z$, regime *(ii)* occurs for $\zeta < -\frac{k}{n}z$ and regime *(iii)* is verified for $\zeta > \frac{n}{k}z$. Regime *(iv)* has no solution since the imaginary part of the impedance should simultaneously verify $\zeta > \frac{n}{k}z$ and $\zeta < -\frac{k}{n}z$. Regime *(iii)* must also be excluded because it violates the admissible phase range of $\varphi_\varepsilon$. Indeed, according to eqs. (3), enforcing $n > 0$ while having $\varphi_\mu > \pi/2$ would necessarily imply a negative $\varphi_\varepsilon$ and therefore $\varepsilon_i < 0$. The unphysical situation is illustrated in Fig.S2.

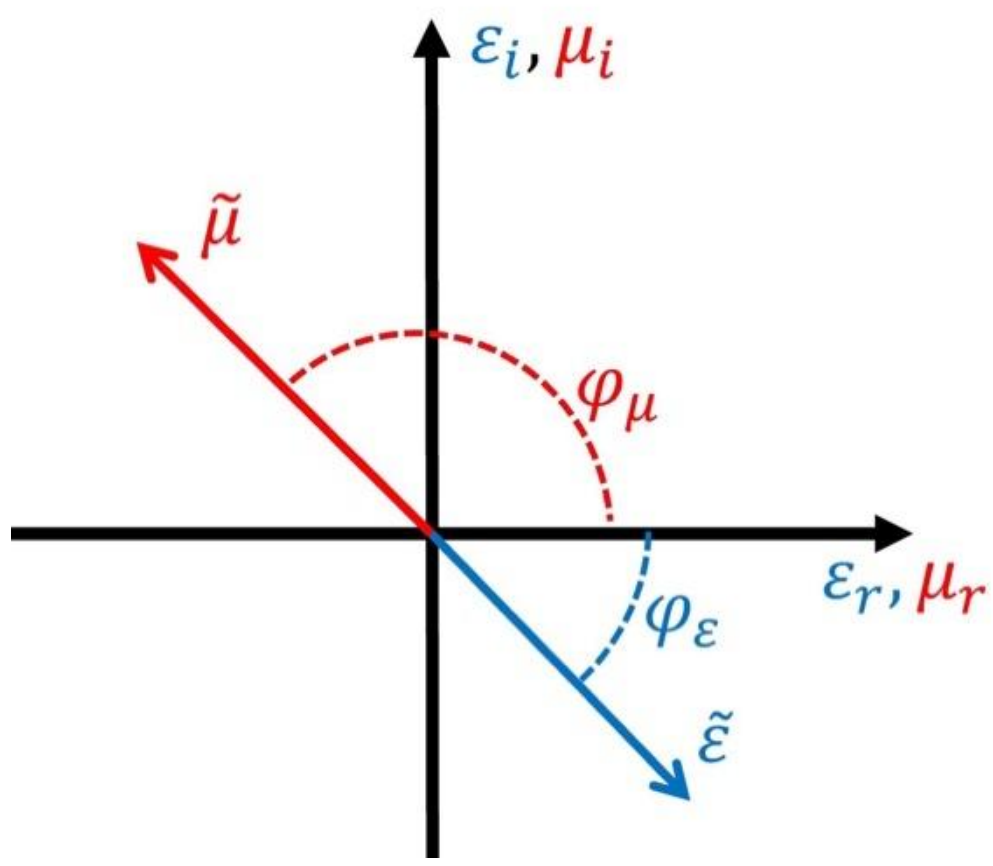

*Figure S2: Polar representations of $\tilde{\varepsilon}$ and $\tilde{\mu}$ to demonstrate that the condition $\mu_r < 0$, $\mu_i < 0$ is incompatible with $\varepsilon_i > 0$.*

Furthermore, if conditions $\varphi_\mu > \pi/2$ and $\varphi_\varepsilon < 0$ are fulfilled, according to eqs. (3), the phase of impedance would satisfy $\varphi_z > \pi/2$ implying $z < 0$.

Once the physical phase range of $\tilde{\mu}$ has been established, it is useful to show that regime *(ii)* corresponds to the $\pi$-shifted configuration of case $\mu_r < 0, \mu_i > 0$.
If the same procedure employed above is instead applied to $\tilde{\varepsilon}$ but assuming $\varphi_\mu \in [0, \pi]$, it results that the only two configurations for the permittivity are: *(j)* $\varepsilon_r > 0$, $\varepsilon_i > 0$ and *(jj)* $\varepsilon_r > 0$, $\varepsilon_i < 0$. Configuration *(jj)* represents the $\pi$-shifted counterpart of the plasma regime. In analogy to the regime (*ii*) of $\tilde{\mu}$, configuration *(jj)* occurs to fulfill $n > 0$ when $\varphi_\mu > \pi/2$.
Previous studies — together with this one — confirm that the solution with $\mu_i < 0$ is valid if $\varepsilon_i > 0$ is kept for any configuration, and vice versa the solution $\varepsilon_i < 0$ is physically possible, provided that $\mu_i > 0$ is verified as universal condition. In [2] the configuration with $\mu_i < 0$ was adopted whereas in [3] and [4] the alternative configuration with $\varepsilon_i < 0$ was employed.

Of course, the negative sign of $\mu_i$ does not imply any violation of the second law of thermodynamics because it belongs to the $\pi$-shifted representation of the case $\mu_r < 0$, $\mu_i > 0$ (regime *(iii)*). Thus, the computation of the dissipated electromagnetic energy density is $\frac{1}{2}(\varepsilon_i E^2 + |\mu_i| H^2)$ independently of the regime of permeability.

## 2. Electrodynamic parameters of the Si substrate

The accurate spectroscopic retrieval of a thin film depends strongly on the reliable acquisition of the substrate impedance as thoroughly discussed in [1]. For this purpose, it is important to apply the Combined Total Variation Technique (CTVT) and retrieve $\tilde{n}$ and $\tilde{z}$, independently.

The CTVT was applied to the intrinsic Si substrates hosting the Al and Cu thin films. The results for a representative sample are reported in Figures S3(a)-(d).

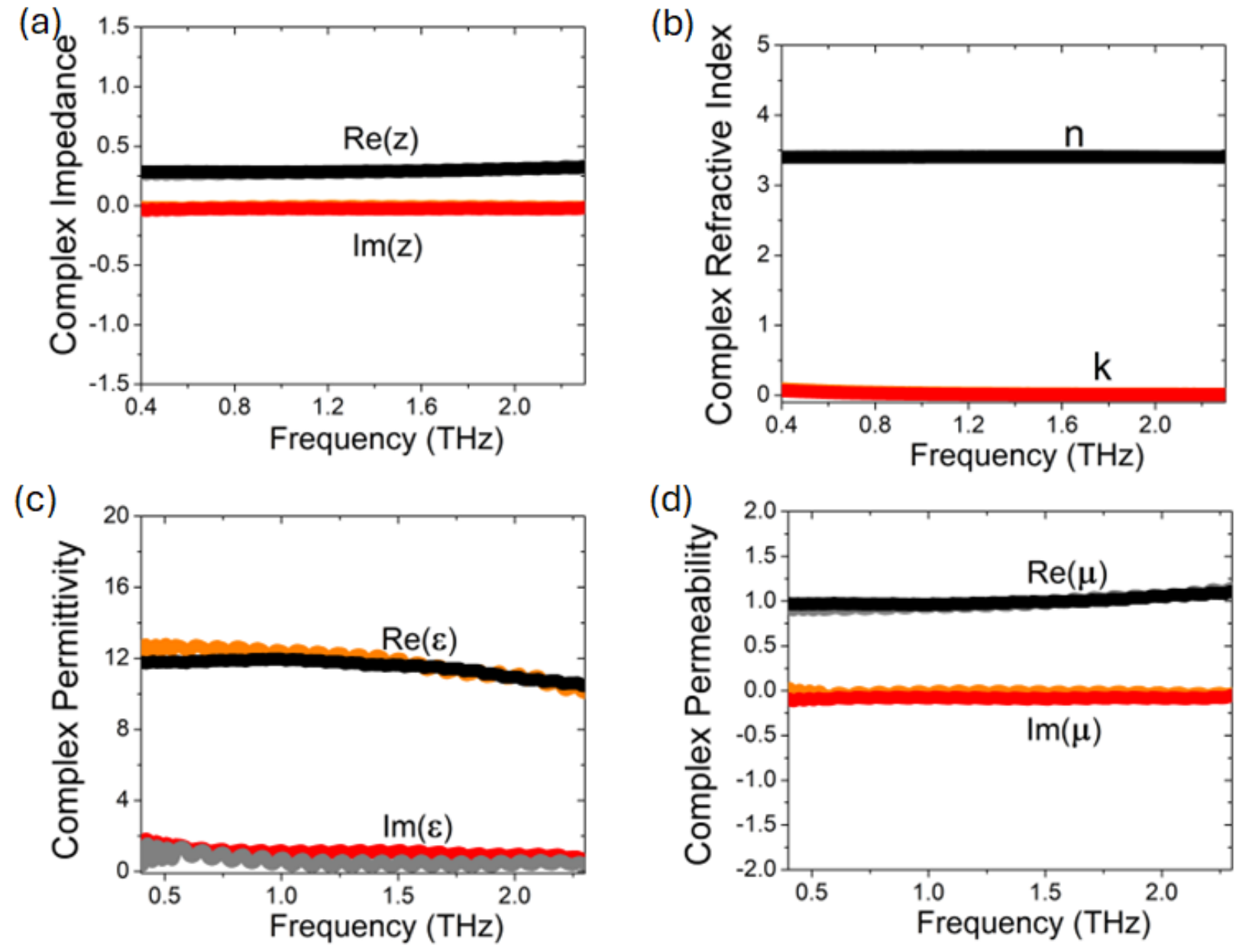

*Figure S3: (a) the complex impedance, (b) the refractive index, (c) the permittivity and (d) the permeability of a Si slab are presented. The color notation follows the same notation used in the paper: black and red for the real and imaginary parts achieved through the minimization of $Err(\tilde{T})$, grey and orange for the corresponding curves obtained through the minimization of $Err(\tilde{R})$.*

The frequency behavior of the electrodynamic complex quantities is accurately determined since the real and imaginary parts achieved from the minimization of $Err(\tilde{T})$ and $Err(\tilde{R})$ functions overlap within the experimental uncertainty.

The resulting impedance $\tilde{z}$ exhibits only a weak frequency dependence up to approximately 2.2 *THz*. Above this frequency, a slight increase in both components becomes noticeable. The imaginary part $Im(\tilde{z})$ remains close to zero, indicating that the impedance is mainly resistive.

The refractive index is substantially non-dispersive across the whole spectral range with $n = 3.42 \pm 0.05$. The extinction coefficient shows a similar behavior $k = 0.04 \pm 0.01$, remaining nearly constant and close to zero, as expected in this frequency region.

The real part of Si permittivity $\varepsilon_r$ is weakly dispersive, spanning the range $[10.5; 12.5]$ whereas the imaginary component $\varepsilon_i$ lies in the interval $[0.8; 1.2]$.

The condition $\tilde{\mu} = 1$ is substantially satisfied.

## References

[1] G. P. Papari, Z. Mazaheri, F. Lo Presti, G. Malandrino, and A. Andreone, Accurate THz measurements of permittivity and permeability of $BiFeO_3$ thin films, Opt. Commun. **586**, 131872 (2025).

[2] D. Schurig, J. J. Mock, and D. R. Smith, Electric-field-coupled resonators for negative permittivity metamaterials, Appl. Phys. Lett. **88**, 1 (2006).

[3] D. R. Smith, D. C. Vier, T. Koschny, and C. M. Soukoulis, Electromagnetic parameter retrieval from inhomogeneous metamaterials, Phys. Rev. E **71**, 036617 (2005).

[4] J. Zhou, L. Zhang, G. Tuttle, T. Koschny, and C. M. Soukoulis, Negative index materials using simple short wire pairs, Phys. Rev. B **73**, 041101(R) (2006).